\documentclass[twocolumn,showkeys,showpacs,preprintnumbers,prd,superscriptaddress,nofootinbib]{revtex4-1}
\usepackage{graphicx}
\usepackage{epsf}
\usepackage{bm}
\usepackage{amsmath}
\usepackage{amsfonts}
\usepackage{amssymb}
\usepackage{epstopdf}
\usepackage{natbib}
\usepackage{hyperref}
\usepackage{color}
\usepackage{verbatim}
\usepackage{multirow}
\usepackage{bm}
\usepackage{hyperref}

\begin{document}

\title{Search for sub-solar mass binaries with Einstein Telescope and Cosmic Explorer}

\author{Rafael C. Nunes}
\email{rafadcnunes@gmail.com}
\affiliation{Divis\~{a}o de Astrof\'{i}sica, Instituto Nacional de Pesquisas Espaciais, Avenida dos Astronautas 1758, S\~{a}o Jos\'{e} dos Campos, 12227-010, S\~{a}o Paulo, Brazil}

\begin{abstract}
A possible detection of sub-solar mass ultra-compact objects would lead to new perspectives on the existence of black holes that are not of astrophysical origin and/or pertain to formation scenarios of exotic ultra-compact objects. Both possibilities open new perspectives for better understanding of our universe. In this work, we investigate the significance of detection of sub-solar mass binaries with components mass in the range: $10^{-2} M_\odot$ up to 1$M_\odot$, within the expected sensitivity of the ground-based gravitational waves detectors of third-generation, viz., the Einstein Telescope (ET) and the Cosmic Explorer (CE). Assuming a minimum of amplitude signal-to-noise ratio for detection, viz., $\rho = 8$, we find that the maximum horizon distances for an ultra-compact binary system with components mass $10^{-2} \, M_\odot$ and 1$M_\odot$ are 40 Mpc and 1.89 Gpc, respectively, for ET, and 125 Mpc and 5.8 Gpc, respectively, for CE. Other cases are also presented in the text. We derive the merger rate, and discuss consequences on the abundances of primordial black hole (PBH), $f_{\rm PBH}$. Considering the entire mass range [$10^{-2}$ - 1]$M_\odot$, we find $f_{\rm PBH} < 0.70$ ($<$ $0.06$) for ET (CE), respectively.   
\end{abstract}



\maketitle

\section{Introduction}
\label{sec:introduction}

We are in the beginning of the era of gravitational wave (GWs) astronomy. The LIGO/VIRGO observatories already detected more than 50 coalescing compact binaries events \cite{theligoscientificcollaboration2021gwtc21, Abbott_2021}, and the probes have targeted binary systems with total masses in the range [2 - 600]$M_\odot$ \cite{Abbott_2016,Abbott_2019}. The LIGO and Virgo detectors are also sensitive to ultra-compact binaries with components below 1$M_\odot$, since the compactness is close to that of the black holes. In the probes for GWs from the coalescence of sub-solar mass binaries, recently performed in \cite{Magee_2018, Abbott_2018, Abbott_Subsolar, Nitz_2021, nitz2021search, Wang_2021, phukon2021hunt}, no convincing candidates were found in LIGO/VIRGO data. 

Search for the sub-solar mass ultra compact binaries is worthwhile because it may provide direct evidence of the existence of black holes that are not of astrophysical origin or formation of exotic ultra-compact objects. We know that in the standard stellar evolution models, the lightest compact objects are formed when stellar remnants exceed the Chandrasekhar mass limit $\sim$ 1.4$M_\odot$. Beyond the Chandrasekhar mass limit, the electron degeneracy pressure in the star's core is insufficient to balance the star's own gravitational self-attraction, and therefore, can no longer prevent the gravitational collapse of a white dwarf. The lightest remnants that exceed the Chandrasekhar mass limit will form neutron stars, and when the neutron degeneracy pressure cannot prevent collapse, heavier stellar remnants will collapse to form black holes. To the present knowledge, there is no model for forming neutron stars for $<$1 $M_\odot$. On the other hand, black holes appear to have a minimum mass $\sim$ 5$M_\odot$. Also, the observations confirm that there is a gap $\sim$ [2, 5]$M_\odot$ between the neutron star and black hole masses  \cite{_zel_2010, Farr_2011, Kreidberg_2012}. Thus, detecting ultra-compact objects below 1$M_\odot$ could challenge the stellar evolution or possibly hint at some unconventional formation scenarios for such objects.

The theoretical postulations for the existence of alternative channels for the formation of black hole were proposed 50 years ago \cite{1967SvA....10..602Z, 1971MNRAS.152...75H, 1975ApJ...201....1C, 1980PhLB...97..383K, 1975Natur.253..251C}. The main motivation is that black holes could have formed in the early universe through the collapse of highly over-dense regions, the so-called primordial black holes (PBHs). It has been shown that PHBs can also form at late times \cite{Bird_2016, Luca_2020}. If PHBs exist, these can naturally account  for the dark matter or at least explain a fraction of the dark matter abundance \cite{Villanueva_Domingo_2021, Carr_2016, Garc_a_Bellido_2017}. Various studies using the observations of black hole mergers by LIGO/Virgo data are carried out to constrain the PBHs and their abundance \cite{Raidal_2017, Wang_2018, Mandic_2016, franciolini2021quantifying, De_Luca_2021, Dom_nech_2021, Khalouei_2021, Wong_2021, Hall_2020, Jedamzik_2021, Lehmann_2020}, including proposals on how to distinguish a PBH from an astrophysical one \cite{Mukherjee_2021, cui2021stochastic}. Also, several analyses and theoretical calculations are carried out to investigate PHBs for the prospects of future GWs detectors, such as LISA \cite{Bartolo_2019,kozaczuk2021signals} and Einstein telescope (ET) \cite{Chen_2020,ng2021singleeventbased}. See \cite{Carr_2020} for a general review on the PBHs. On the other hand, various proposals for non-baryonic dark matter models can produce subsolar mass black holes, as well as possibilities for the formation of some exotic ultra-compact objects, with masses below 1$M_\odot$ \cite{Cardoso_2019}. Thus,  the detection of sub-solar mass ultracompact objects would provide the cleanest signature of such scenarios.

The aim of this work is to search for the possible imprints of sub-solar mass binaries within the expected sensitivity of Einstein Telescope (ET) and Cosmic Explorer (CE). Both instruments are ground-based GWs detectors of third-generation, which could be operating in the mid 2030s. With ET and CE, we will be able to determine the nature of the densest matter in the universe; reveal the universe’s binary black hole and neutron star populations throughout cosmic time; provide an independent probe of the history of the expanding universe; physics near the black hole horizon; testing exotic compact objects, as well as many other questions in fundamental physics and cosmology. See \cite{Maggiore_2020, sathyaprakash2019cosmology, reitze2019cosmic} for a presentation of scientific objectives with these observatories. In this paper, we show that ET and CE will be able to detect strong signals coming from sub-solar mass binaries system candidate with components mass in the range $\in [10^{-2} - 1.0]$ $M_\odot$. Estimating the merger rate of these compact binaries, we discuss consequences on the PBHs' abundance. 

This paper is structured as follows. In next section, we define the essential quantities to analyse the GWs signals. In Section \ref{sec:results}, we present our main results and lastly, in Section \ref{sec:conclusions}, we outline our final considerations and perspectives.
 
\section{Analysis strategy}
\label{sec:Analysis}

In this section, we briefly summarize the methodology and main information used to search for compact binary system. For a given GW strain signal $h(t) = A(t) \cos [\Phi(t)]$, one can use the stationary-phase approximation for the orbital phase of inspiraling binary system to obtain its Fourier transform $\tilde{h}(f)$. In the case of a coalescing binary system, we have
\begin{equation}
\label{waveform}
\tilde{h}(f) = Q \mathcal{A} f^{-7/6} e^{i\Phi(f)}\ ,
\end{equation}
where $\mathcal{A}$ is the GW amplitude computed perturbatively within the so-called post-Newtonian formalism (PN), and can be written as
\begin{equation}
\label{A}
\mathcal{A} =  \sqrt{\dfrac{5}{96}} \dfrac{\mathcal{M}^{5/6}_c}{\pi^{2/3} d_L} \left( \sum_{i=0}^6 A_i(\pi f)^{i/3} \right).
\end{equation}
Here $d_L$ is the luminosity distance, and the function $Q$ is given by
\begin{equation}
\label{Q}
Q^2 = F^2_{+}(1+cos^2(\iota))^2 + 2F^2_{\times}cos^2(\iota),
\end{equation}
where $\iota$ is the inclination angle of the binary orbital angular momentum with respect to the line of sight, and $F^2_{+}$, $F^2_{\times}$ are the pattern functions (specific functions for each detector).

In Eq.~(\ref{waveform}), the function $\Phi(f)$ is the inspiral phase of the binary system:
\begin{equation}
\label{phi}
\Phi(f) = 2 \pi f t_c - \phi_c - \dfrac{\pi}{4} + \dfrac{3}{128 \eta v^5} \left[ 1 + \sum_{i=2}^7 \alpha_i v^i  \right],
\end{equation}
where the coefficients $\alpha_i$ are computed perturbatively in a post-Newtonian formalism.

In this work, we will use the TaylorF2 waveform model, which uses the stationary phase approximation for the waveform, and the 3.5 PN expression for the orbital phase of inspiraling binary black holes with aligned spins, along with the tidal effects on the phase (up to the 6PN level) for compact objects. In the above equation, we have defined $v \equiv (\pi M f)^{1/3}$, $M \equiv m_1 + m_2$, $\eta \equiv m_1 m_2/(m_1 + m_2)^2$, and $\mathcal{M}_c \equiv M \eta^{3/5}$ to be the inspiral reduced frequency, total mass, symmetric mass ratio, and the chirp mass, respectively. The quantities $t_c$ and $\phi_c$ are the time and phase of coalescence, respectively. 

The amplitude of signal-to-noise ratio (SNR), $\rho$, for a deterministic signal $\tilde{h}(f)$ is given by

\begin{equation}
\rho^2 \equiv 4 {\rm Re} \int_{f_{low}}^{f_{upper}} \, \dfrac{\vert \tilde{h}(f)\vert ^2}{S_n} df,
\label{eq:snr}
\end{equation}
where $S_n(f)$ is the detector spectral noise density.
\\

\begin{figure}
\begin{center}
\includegraphics[width=3.3in]{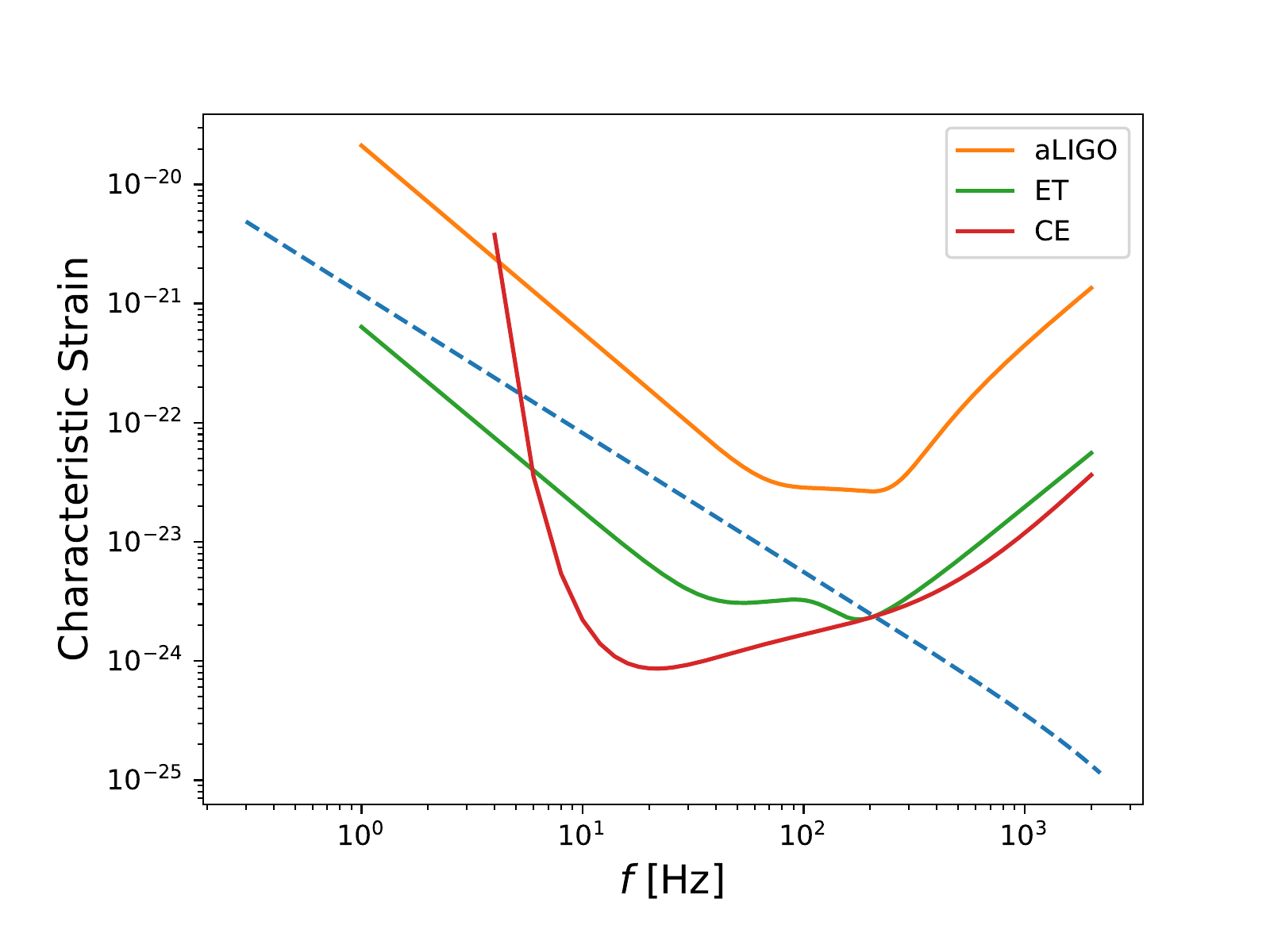} 
\caption{Characteristic strain of a possible sub-solar mass binary system candidate with components mass 1$M_\odot$ at $d_L =$ 100 Mpc plotted along with $\sqrt{f S_n(f)}$, where $S_n(f)$ is the noise power spectral density for ET, CE and aLIGO.}
\label{PSD}
\end{center}
\end{figure}

{\it Sensitivity curves}: We considered the ET and CE power spectral density noises. Both instruments are third-generation ground detectors, covering the frequency range 1-$10^{4}$ Hz. The signal amplitude sensitivity of ET and CE is expected to be more than ten times larger than the current advanced ground-based detectors. For ET, we consider the ET-D sensitivity curve \cite{ET,Hild_2011,Maggiore_2020}. For CE, we also consider the amplitude spectrum of the detector noise also publicly available in \cite{CE,sathyaprakash2019cosmology}.

Figure \ref{PSD} shows the characteristic strain for a qualitative example considering a sub-solar mass binary system with components mass 1$M_\odot$ at $d_L = 100$ Mpc, along with the noise power spectral density for ET, CE and aLIGO. In all the results presented in this work, we use the integration approach on the average over all possible directions and inclinations.

Very low mass systems are expected to emit GWs in the frequency range of the ground-based detectors. To understand it, to leading post-Newtonian order, the frequency as a function of time is given by $f(t) = \frac{1}{8 \pi G M_c/c^3} \Big( \frac{5 G M_c/c^3}{t-t_c} \Big)^{3/8}$. For example, an equal mass binary with components masses 1$M_\odot$, 0.2$M_\odot$, will have GW frequencies of 0.25Hz, 0.70Hz, respectively, one year prior to merger. The maximum frequency can be determined by the frequency of the innermost stable circular orbit ($f_{\rm ISCO}$), where $f_{\rm ISCO} = \frac{c^3}{6 \sqrt{6} \pi G M}$. For binary system with components masses 1$M_\odot$, 0.2$M_\odot$, we have $f_{\rm ISCO} =$2200 Hz, 10100 Hz, respectively. Thus, sub-solar mass binaries compact inspiraling can be visible at the maximum frequency range sensitivity of the ground-based detectors before the merger. On the other hand, even considering several years prior to merger, the expected GW amplitude will be beyond and below the LISA band operation. These qualitative aspects are clear in Figure \ref{PSD}. We assume that the coalescence of sub-solar mass black hole binaries have negligible spin. This is consistent with the predictions of spin distributions presented in \cite{Chiba_2017, Luca_2019, Postnov_2019, Luca_2020}.
\\

{\it Merger rate for null results}:
We can calculate the maximum distance for which an optimally located and oriented source would be observed with some $\rho$ value. In general, the detectors will measure a weaker response to GWs, depending on the location and orientation of the binary system.
This reduction is quantified through the antenna patterns, $F_+$ and $F_\times$, which always take values less than 1. As demonstrated in \cite{Finn_1993,Abadie_2010}, after averaging the detector response over both location and orientation, the binary system will reduce the strain recovered by a factor of 2.26. Thus, this can be used to define the average range of the detector as
\begin{align} D_{avg} = \frac{D_{max}}{2.26}. \end{align}

The average sensitive distance allows us to approximate limits on the
coalescence rate from null results for a general GWs search. The
loudest event statistic formalism~\cite{Biswas_2009} states that we can
constrain the binary merger rate for a specific mass bin, $i$, at $90\%$
confidence level (CL) as
\begin{align}
\mathcal{R}_{90, i} = \frac{2.3}{\langle VT \rangle_i},
\end{align}
where $\langle VT \rangle_i$ is the sensitive volume-time, and is given by 

\begin{align} 
\langle VT \rangle_i = \tfrac{4}{3} \pi D_{avg, i}^3 T. 
\end{align}
Here $T$ is the analyzable live-time of the detectors. This method provides an excellent approximation of the sensitive 4-volume. We will use this methodology to estimate the rates in the sub-solar mass region. Similar approach has been applied previously in \cite{Magee_2018}. We assume $T = 1$ yr, in all our results.

\begin{figure*}
\begin{center}
\includegraphics[width=3.3in]{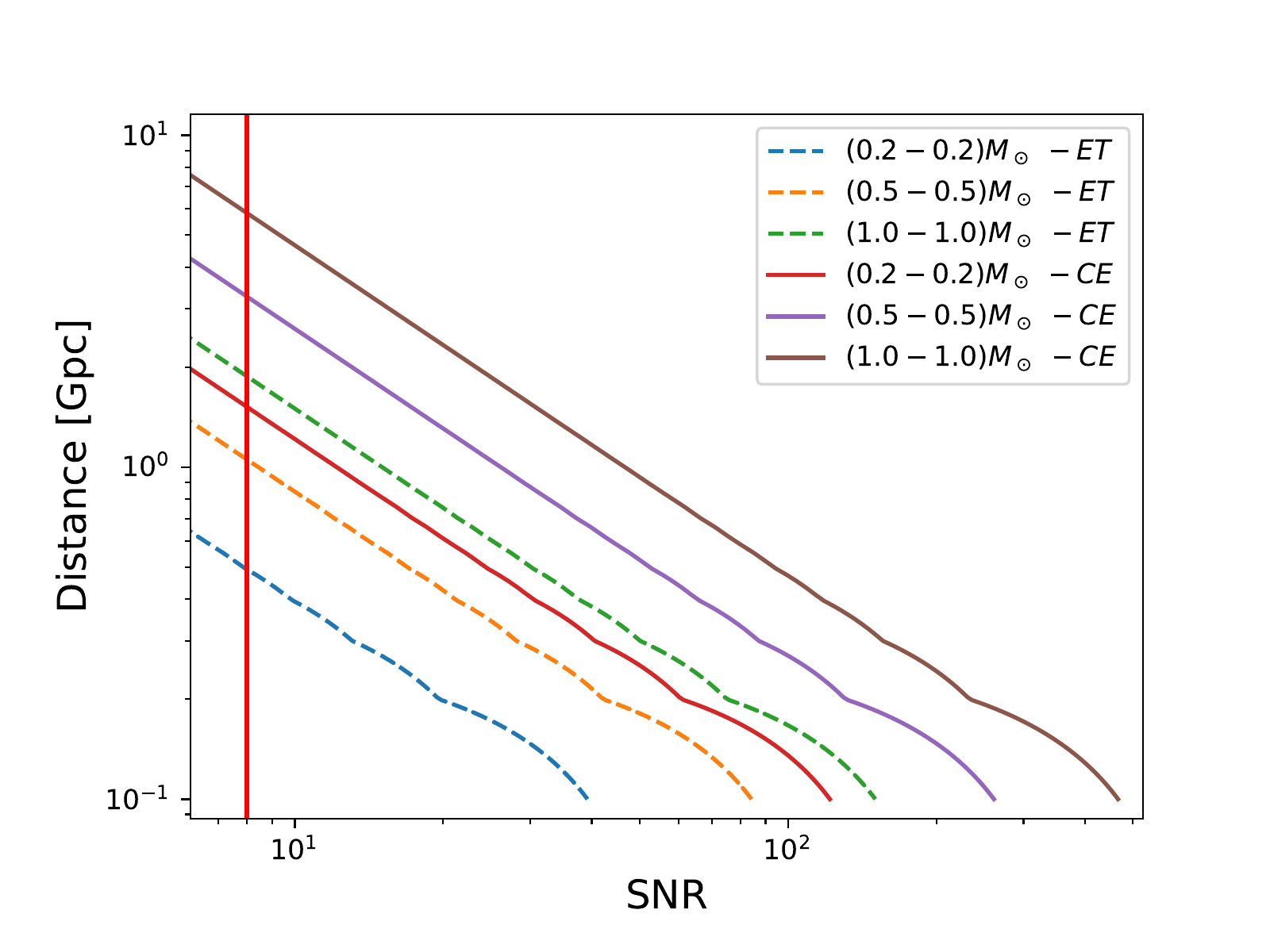} 
\includegraphics[width=3.3in]{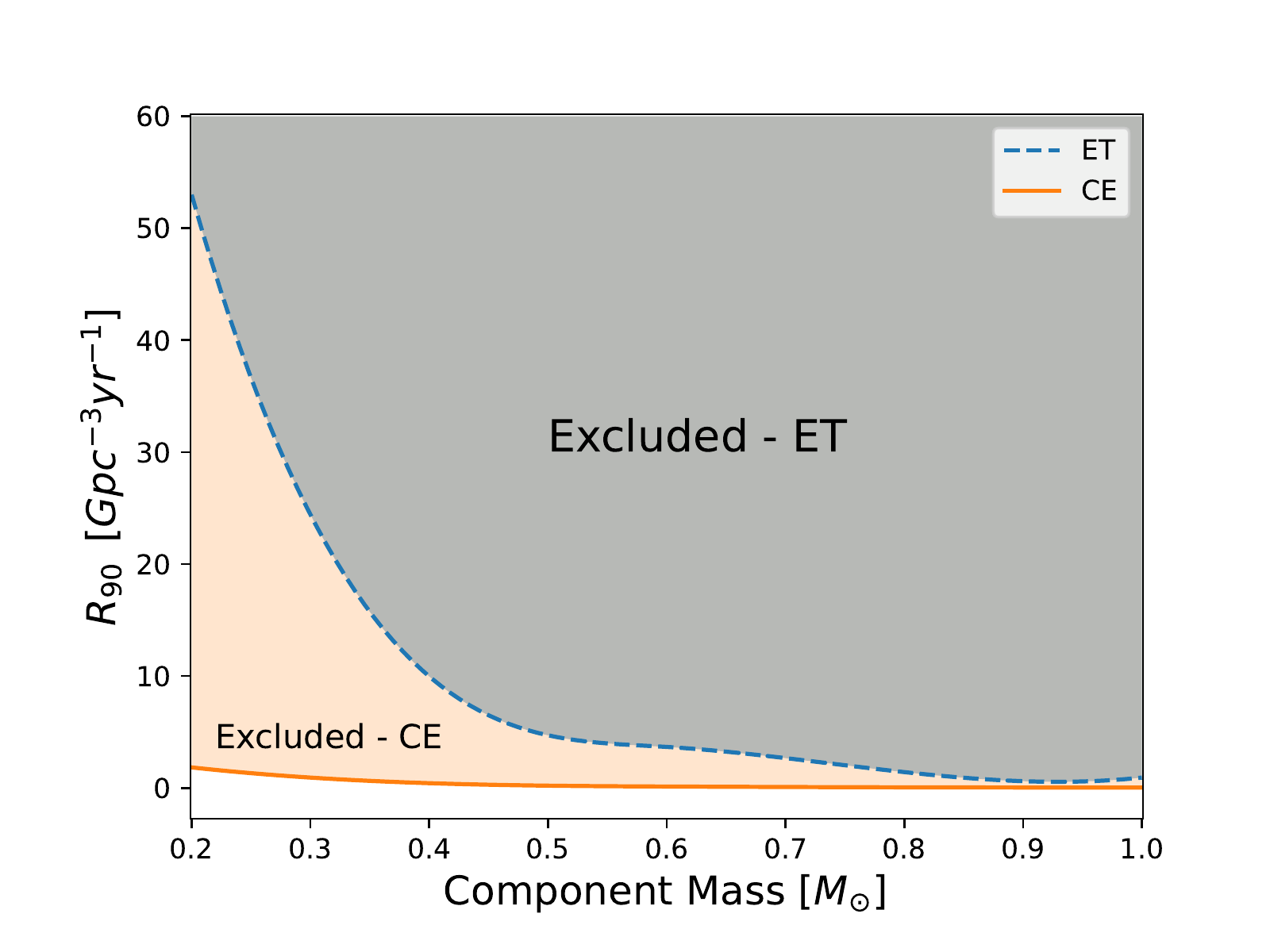} 
\caption{Left panel: The distance in Gpc units to an optimally oriented, equal mass binary as a function of the signal-to-noise ratio (SNR) for the perspective of the  ET and CE sensitivity noise curve. The vertical line represent $SNR \equiv \rho = 8$. Right panel: The merger rate of equal-mass ultra-compact binaries as a function of the components mass in units of $M_\odot$.}
\label{Fig1}
\end{center}
\end{figure*}

\begin{table}
    \centering
    \caption{Estimates of the merger rate of equal-mass compact binaries in the range [0.2, 1.0]$M_\odot$ for ET and CE instruments.}
    \label{tab1}
    \begin{tabular}{ccccccc}
        \hline
        \hline
		Instrument & Component Mass $[M_\odot]$ & $\mathcal{R}_{90} \,\, [{\rm Gpc}^{-3} {\rm yr}^{-1}]$  \\ 
		\hline
		ET & 0.2 & 53 &  \\ 
		ET & 0.5 & 4.7 &  \\  
		ET & 1.0 & 0.924 &  \\
		CE & 0.2 & 1.829 &  \\ 
		CE & 0.5 & 0.090 &  \\  
		CE & 1.0 & 0.029 &  \\
		\hline
    \end{tabular}
\end{table}

\section{Results}
\label{sec:results}

Figure \ref{Fig1} on the left panel shows the distance (horizon distance) to an optimally oriented, and some equal mass binary in the range [0.2 - 1.0]$M_\odot$ as a function of the SNR obtained using the ET and CE power spectral density noises. The vertical line represents $\rho = 8$. Within the perspective of ET sensitivity, we note that the maximum distance for detection, assuming the reference value $\rho = 8$, is 0.47 Gpc, 1.05 Gpc and 1.89 Gpc for compact binaries with equal components mass 0.2$M_\odot$, 0.5$M_\odot$ and 1.0$M_\odot$, respectively. For CE sensitivity, we find 1.5 Gpc, 32. Gpc and 5.8 Gpc for equal components mass with 0.2$M_\odot$, 0.5$M_\odot$ and 1.0$M_\odot$, respectively. Any other combination between these masses will generate intermediate results to these. In these simulations, we consider $f_{low} = 10$Hz and $f_{upper} = f_{ISCO}$. Evidently, the horizon distance for CE is greater than ET, because CE has a greater sensitivity.

Figure \ref{Fig1} on the right panel shows the constraints on the merger rate of equal-mass ultra-compact binaries in the range [0.2, 1.0]$M_\odot$ for both, ET and CE instruments. Table \ref{tab1} summarizes the estimates for some particular cases. The results for CE can improve up to 1 order of magnitude of the expected merger rate for ET. We do not take into account possible eccentric orbits effects, which may possibly increase the expected value for $\mathcal{R}_{90}$ \cite{Wang_2021,nitz2021search}.
\\

\subsection{Bounds on primordial black holes}

There is a strong theoretical appeal for the existence of PBHs, especially because PBHs are dark matter candidates in a broad mass range. Constraint on the binary merger rate places bounds on the total fraction of dark matter made of PBHs, which can be quantified by the parameter $f_{\rm PBH}$. Several authors have shown possible ways to form PBHs in the early Universe \cite{1967SvA....10..602Z, 1971MNRAS.152...75H, 1975ApJ...201....1C, 1980PhLB...97..383K, Clesse_2017, Sasaki_2018, Raidal_2017} and sub-solar mass PBHs are proposed to exist in various scenarios, for instance, see \cite{Nakamura_1997, Ioka_1998, Carr_2021}. From the LIGO/Virgo data, constraints are achieved on sub-solar sources to $< 1.0 \times 10^6 {\rm Gpc}^{-3} {\rm yr}^{-1}$ and $< 1.9 \times 10^4 {\rm Gpc}^{-3} {\rm yr}^{-1}$ for (0.2 $M_\odot$, 0.2$M_\odot$) and (1.0 $M_\odot$, 1.0$M_\odot$) ultra-compact binaries \cite{Abbott_2018}. Other analyses to search for sub-solar mass compact-binary mergers in LIGO/VIRGO data that constrain the PBHs populations are presented in \cite{Magee_2018, Nitz_2021, nitz2021search, Wang_2021, phukon2021hunt}. In particular, see \cite{phukon2021hunt} for a summary of the search for non-spinning binary sources, spanning sub-solar mass ranges.

On the other hand, the merger rate of the sub-solar GWs sources is model-dependent, and can depend on different formation scenarios \cite{Luca_2020_2, Jedamzik_2020, Ali_Ha_moud_2017, Gow_2020}. The merger rate assuming a Poisson scenario can be written as 

\begin{align}
\begin{split}\label{mergerpbh}
    \frac{R_{PBH}(t)}{\text{Gpc$^{-3}$ yr$^{-1}$}}= 1.6\times 10^6f_{\text{sup}}f_{\text{PBH}}^{53/37}& \eta^{-34/37}\bigg(\frac{M}{M_\odot}\bigg)^{-32/37}\\ & \times \bigg(\frac{t}{t_0}\bigg)^{-34/37},
    \end{split}
\end{align}
where $f_{\text{sup}}$ is the suppression factor which varies from $10^{-3}$ to 1 \cite{Raidal_2019}; $f_{\text{PBH}}$ is the fraction of PBHs in dark matter; $\eta$ denotes the reduced mass ratio; $M$ denotes the total mass; $t$ denotes the proper time, and $t_0$ denotes the age of the universe today.

\begin{table}
    \centering
    \caption{Upper bounds on $f_{\text{sup}}$ and $f_{\text{PBH}}$ (fraction of PBHs in dark matter) derived from the expected event rate within the ET sensitivity.}
    \label{tab:2}
    \begin{tabular}{ccc}
        \hline
		Component Mass $[M_\odot]$ & $f_{sup}$ & $f_{PBH}$ \\ 
		\hline
		0.2 & $< 0.47$ & $< 0.022$ \\ 
		0.5 & $< 0.43$ &  $< 7 \times 10^{-3}$ \\ 
		1.0 & $< 0.17$ &  $< 3.4 \times 10^{-3}$ \\ 
		\hline
    \end{tabular}
\end{table}

\begin{figure*}
\begin{center}
\includegraphics[width=3.in]{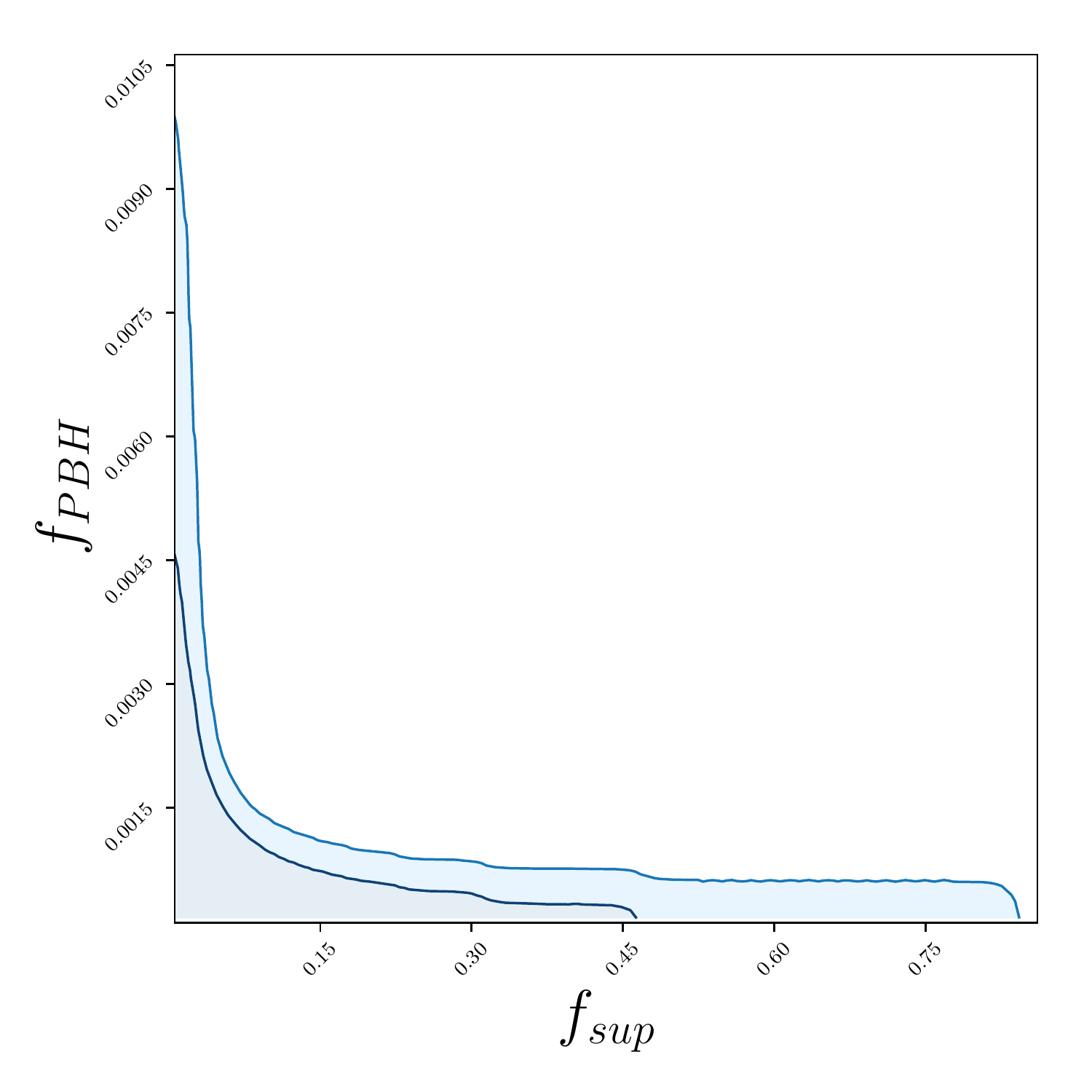} 
\includegraphics[width=3.in]{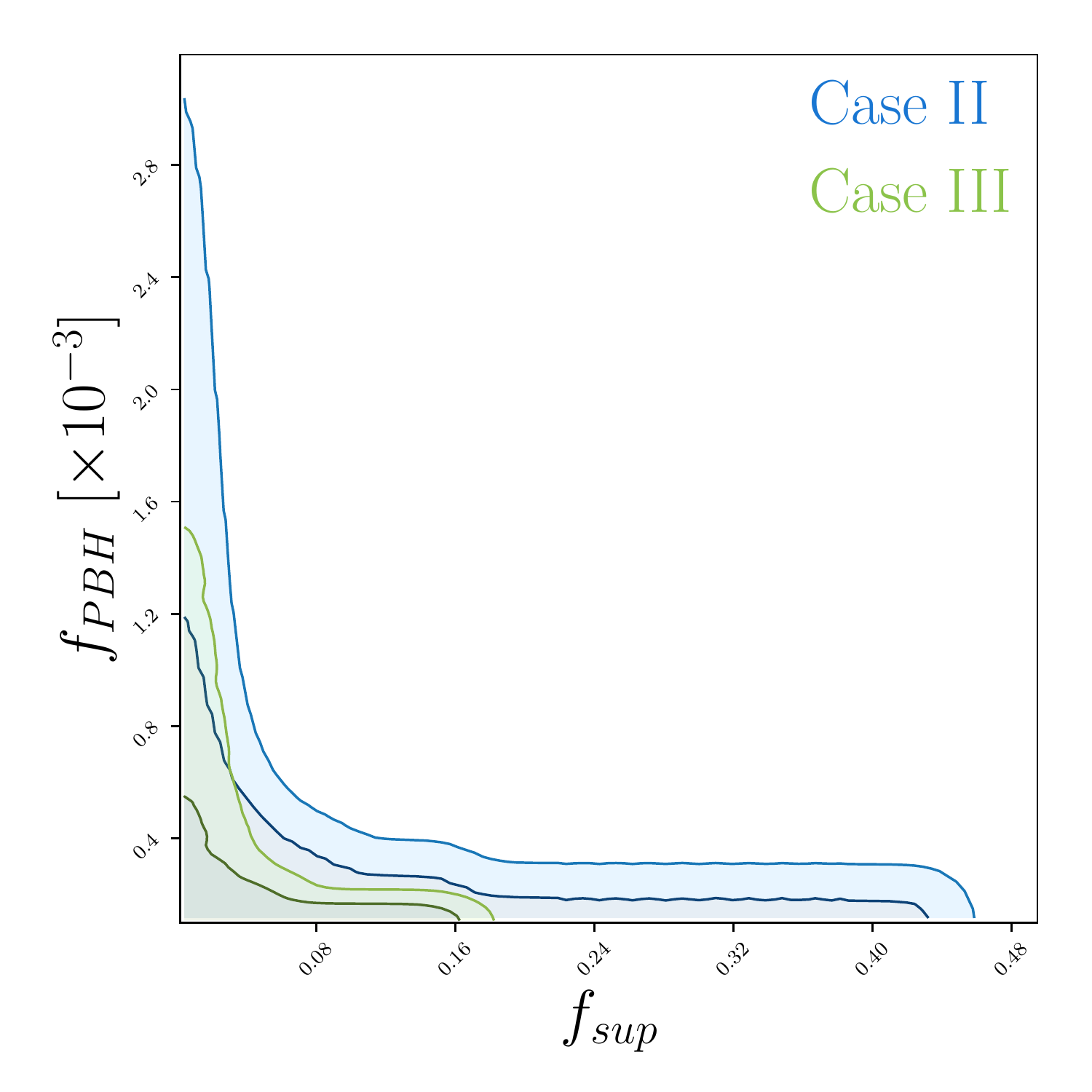} 
\caption{Left panel: Two-dimensional joint posterior distributions in the $f_{sup}$ - $f_{PBH}$ plane, with the corresponding 68\% CL and 95\% CL contours, obtained from the expected event rate within the ET sensitivity assuming a binary system with component mass 0.2$M_\odot$. Right panel: Same as in left panel, but for a binary system with equal components mass 0.5$M_\odot$ (Case II) and 1.0$M_\odot$ (Case III).}
\label{PS}
\end{center}
\end{figure*}

We use the MCMC method to analyze the parameters ${\theta_i}  = \Big\{ f_{\text{sup}}, f_{\text{PBH}} \Big\}$, building  the posterior probability distribution function: 
\begin{equation}
\label{psd}
p(\theta_i, \alpha|D) = \frac{1}{Z} p(\theta,\alpha) p(D|\theta,\alpha)  \, ,
\end{equation}
where $p(\theta,\alpha)$ and $p(D|\theta,\alpha)$ are the prior distribution and the likelihood function, respectively. Here, the quantities $D$ and $\alpha$ are the set of observations and possible nuisance parameters. $Z$ is a normalization term. We perform the statistical analysis based on the \textit{emcee} algorithm \cite{Foreman_Mackey_2013}, assuming the theoretical $R_{PBH}$ model described above with the following uniform priors on the parameters: 
$f_{\text{sup}} \in [10^{-3}, 1]$ and $f_{\text{PBH}} \in [0, 1]$. During our analysis, we discarded the first 20\% steps of the chain as burn-in.

Table \ref{tab:2} shows the upper bounds on $f_{\text{sup}}$ and $f_{\text{sup}}$ derived from the expected event rate within the ET sensitivity assuming component mass with 0.2$M_\odot$, 0.5$M_\odot$ and 1.0$M_\odot$. Estimates based on CE are one order of magnitude smaller than these. Figure \ref{PS} on the left panel shows the parametric space limited to 68\% CL and 95\% CL for the case with component mass 0.2$M_\odot$. On the right panel, we show the case with component mass 0.5$M_\odot$ (label Case I) and 1.0$M_\odot$ (label Case II).

It shows that matching the constraints derived from these three mass bins, we can explain $\sim$ 2.3\% of total dark matter abundance from the ET sensitivity. Using CE, we note $\sim$0.2\% of total dark matter abundance. 
\\

\begin{figure}
\begin{center}
\includegraphics[width=3.in]{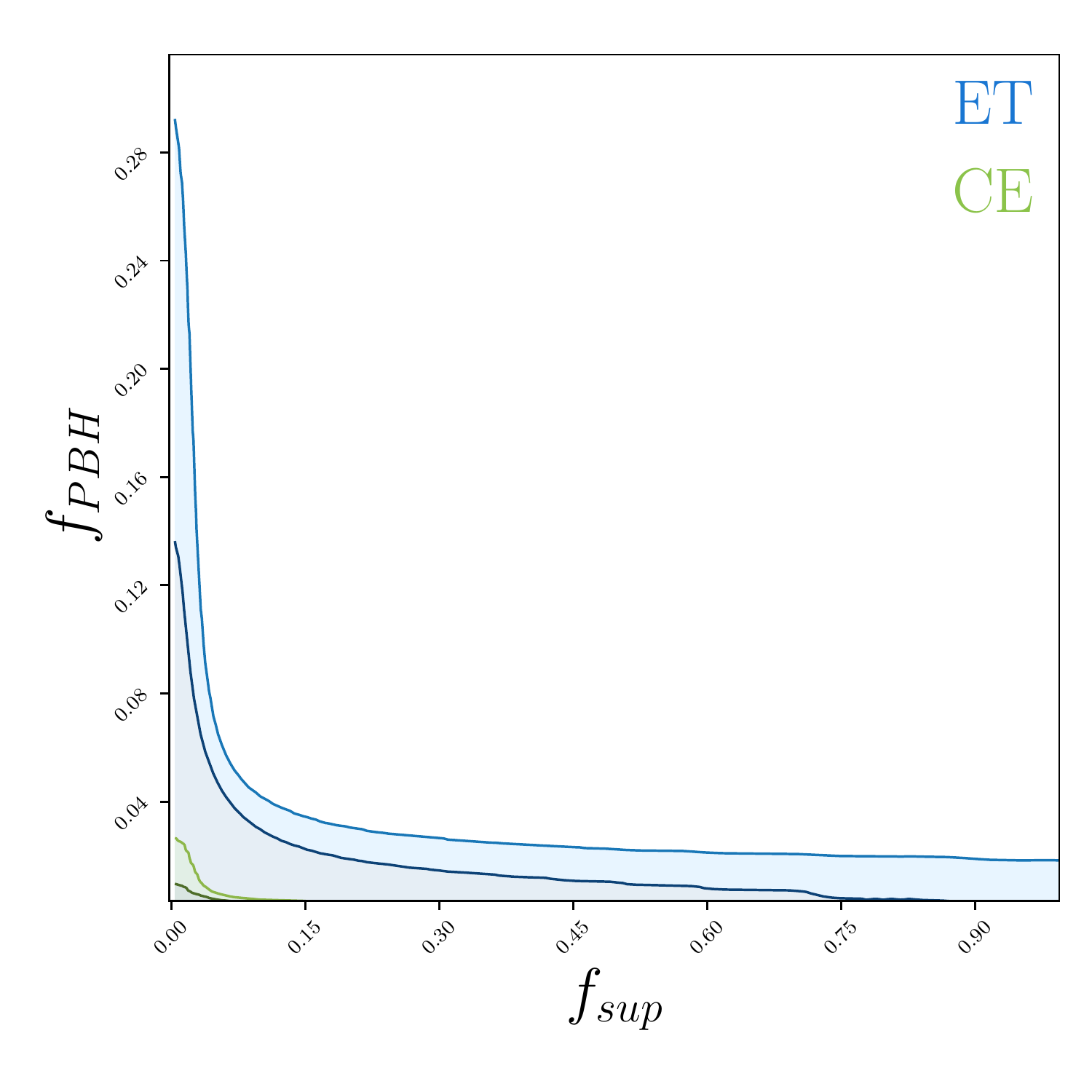} 
\caption{Two-dimensional joint posterior distributions in the $f_{sup}$ - $f_{PBH}$ plane, with the corresponding 68\% CL and 95\% CL contours, obtained from the expected event rate within the ET and CE sensitivity assuming a binary system with component mass 0.01$M_\odot$.}
\label{PS2}
\end{center}
\end{figure}

{\it Ultra-compact binary system with component mass 0.01$M_\odot$}: We repeat the same analysis strategy, but now to verify the feasibility of detecting an ultra-compact binary system with component mass 0.01$M_\odot$. Assuming a minimum $\rho= 8$, we find that the maximum horizon distances are $\sim$40 Mpc and $\sim$125 Mpc for ET and CE, respectively. The estimates of the merger rate of this compact binaries are 99034 ${\rm Gpc}^{-3} {\rm yr}^{-1}$ and 3245 ${\rm Gpc}^{-3} {\rm yr}^{-1}$ for ET and CE, respectively. The general upper bounds on $f_{\text{PBH}}$, which can fit these merger rates are $f_{\text{PBH}} < 0.68$ and $f_{\text{PBH}} < 0.06$ for ET and CE, respectively. We show the parametric space limited to 68\% CL and 95\% CL in Fig. \ref{PS2}.

Therefore, within ET perspective, binary system with component mass 0.01$M_\odot$, can represent up to $\sim$68\% of dark matter (if $f_{sup} = 10^{-3}$). Taking the contributions across the mass range [$10^{-2}$, 1.0]$M_\odot$ combined, we have $f_{PBH} < 0.70$, that is, a limit of $\sim$ 70\%. These constraints can be significantly improved within CE perspectives, where we note $f_{PBH} < 0.06$, finding a maximum $\sim$ 6\% for the abundance of dark matter.

\subsection{Other alternatives for sub-solar-mass objects}

It is now clear that it will be possible to detect sub-solar-mass objects with high significance (high $\rho$ value), from 40 Mpc up to a few Gpc distance with ET and CE. See Figure \ref{Fig1} on the left panel for a summary. In addition to interpreting these ultracompact objects as PHBs, there is a wide range of theoretical predictions, which in principle, lead to the formation of objects with mass below 1$M_\odot$.

In \cite{Kouvaris_2018}, the authors proposed a mechanism that can convert a sizeable fraction of neutron stars into BHs with mass $\sim$ 1$M_\odot$, too light to be produced via standard stellar evolution. Such BHs could be in binary systems, and thus may be searched by GWs detectors. Also, sub-Chandrasekhar mass BHs are also demonstrated to exist in \cite{Dasgupta_2021}, where stellar objects catastrophically accrete non-annihilating dark matter, and the small dark core subsequently collapses, eating up the host star and transmuting it into a BH. Rotating dark stars, constituted for both, fermionic and bosonic equations of state, in the presence of self-interacting dark matter, can also generate ultra-compact objects with $<$ 1$M_\odot$ (see \cite{Maselli_2017} and references therein). Quark stars \cite{Cardoso_2019}, anisotropic dark matter stars \cite{Moraes_2021}, and many other mechanisms can form sub-solar-mass objects (see \cite{Cardoso_2019} for a review). Therefore, there is a rich source of models and physics with sub-Chandrasekhar mass, which can not be explained by stellar evolution. These may certainly involve a new physics, and may be alternatives to PBHs. Certainly the mechanism generation of GWs in these systems must be better modeled and understood, to search possible imprints of the such ultracompact objects.
\\

\section{Final remarks}
\label{sec:conclusions}

We have presented the search (a forecast) for ultracompact binary mergers with components mass below 1$M_\odot$ within the expected sensitivity for the ET and CE detectors. We have concluded that ultracompact binary systems with equal component mass of $10^{-2}$ $M_\odot$ up to 1$M_\odot$ could be detected with high significance since 40 Mpc (125 Mpc), for very low mass system with $10^{-2}$ $M_\odot$ in the ET (CE) sensitivity, up to 1.89 Gpc (5.8 Gpc), for binary system with component mass 1$M_\odot$, in the ET (CE) sensitivity band, respectively. For possible components mass of the order of magnitude less than $10^{-2}$, within the approach developed here, it will be difficult to have  signals with significant SNR values, that is, $\rho > 8$. 

We have determined the merger rate in the mass range [$10^{-2}$ - 1]$M_\odot$, and then constrained the abundance of primordial black holes as a fraction of the total
dark matter in this range mass, quantifying $f_{\text{PBH}} < 0.70$ and $f_{\text{PBH}} < 0.06$, from the perspective of ET and CE, respectively. Therefore, CE puts tight constraint on $f_{\text{PBH}}$. Considering non-negligible spin, eccentric orbits, and possible tidal deformability effects on the waveform can improve and bring new perspectives in this regard, since the origin of these systems can have very different physical aspects. 
On the other hand, still in this generation of observations, the Advanced LIGO/Virgo in their final design sensitivities, will be more sensitive to detect possible mergers of ultracompact binaries, which may open new trends for new physics involving sub-solar mass ultracompact objects.

\begin{acknowledgments}
\noindent  

I am grateful to Jose C. N. de Araujo and Suresh Kumar for very constructive comments and suggestions. Also, I would like to thank the financial support from the Funda\c{c}\~{a}o de Amparo \`{a} Pesquisa do Estado de S\~{a}o Paulo (FAPESP, S\~{a}o Paulo Research Foundation) under the project No. 2018/18036-5.

\end{acknowledgments}

\bibliography{Ref}

\end{document}